# The Benefit of Wide Energy Range Spectrum Acquisition During Sputter Depth Profile Measurements

Uwe Scheithauer, 82008 Unterhaching, Germany
E-Mail: scht.uhg@googlemail.com
Internet: orcid.org/0000-0002-4776-0678; www.researchgate.net/profile/Uwe_Scheithauer

**Abstract:**

Thin film systems are often analysed by using sputter depth profiling. First the sample gets eroded by inert gas ion impact during sputter depth profiling. Then the elemental composition of the freshly unveiled surface is determined by using a surface sensitive analytical method as AES or XPS, for instance. This way the depth distributions of the elements are recorded as a function of sputter time. The time to record the spectral data in a certain sputter depth is kept as short as possible to avoid recontamination of the freshly sputtered surface by adsorption of gas particles from the vacuum during the measurement. Therefore in every sputter depth only those spectral regions are recorded, which belong to the elements expected to be in the sample.

But in case the sample composition is entirely unknown, it is indispensable to measure wide energy range spectra. By this approach the depth distributions of all elements are estimated, which are detectable by the used analytical method. The measurement of wide energy range spectra during sputter depth profile acquisition is promising for samples, which are very insulating. If the surface potential varies in different sputter depths, the elemental peaks are shifted in an unpredictable way. If wide energy range spectra are recorded, the energy scale can be re-adjusted afterwards and the data can be re-evaluated. Both use cases are demonstrated exemplarily by sputter depth profile measurements.

___________________________________________

## 1. Introduction

In-depth profiling of thin film systems is one of the important applications of Auger electron spectroscopy (AES) and X-ray photoelectron spectroscopy (XPS) for materials analysis purposes. Applying this technique the sample surface gets eroded by ion bombardment ("sputtering") usually using inert gas ions in the energy range between 250 eV and 5 keV. The residual surface is analyzed after each sputter step. The depth distributions of the elements are recorded as a function of sputter time [1, 2]. To avoid recontaminations of the sputtered surface by adsorption of gas particles from the residual vacuum, the measurement time is chosen as short as possible. Therefore in each sputter depth only the small energy regions of the spectrum belonging to the elements of interest are measured to minimise misleading recontaminations.

But for some analytical tasks it is worth to measure a wide energy range spectrum. This analysis approach works for a sample of completely unknown compositions. In this case the depth profile analysis starts without any assumption concerning the elemental composition of the sample or its layer structure. If a wide range energy spectrum is measured, all information attainable by the analysis method is recorded. In a retrospective approach the data can be re-evaluated without serious limitations. In the first example the analysis of a contamination layer of unknown composition [3] demonstrates the benefit of wide area spectrum acquisition during sputter depth profile measurement. A second depth profile is measured on a thin film system sample, which shows large shifts of the spectra because some layers are very insulating. The surface potential varies in different sputter depths, the elemental peaks are shifted in an unpredictable way. But since wide energy range spectra are recorded, the energy scale can be readjusted after the measurement and the data can be re-evaluated.

## 2. Instrumentation

The XPS measurements were done using a Physical Electronics XPS Quantum 2000. This XPS microprobe achieves its spatial resolution by the combination of a fine-focused electron beam generating the X-rays on a water cooled Al anode and an elliptical mirror quartz monochromator, which monochromatizes and refocuses the X-rays to the sample surface. Details of the instruments design and performance are discussed in literature [4-10]. For sputter depth profiling the instrument is equipped with a differentially pumped $Ar^+$ ion gun. Sputter ion energies are selectable between 250 eV and 5 keV. For a flat mounted sample as used here in a Quantum 2000 the incoming X-rays are parallel to the surface normal. In this geometrical situation, the mean geometrical energy analyser take off axis and the differentially pumped $Ar^+$ ion gun are oriented ~45° relative to the sample surface normal. Data evaluation was done by an improved version of the PHI software MultiPak 6.1 [11]. It has been improved by a module, which enables an evaluation of measured depth profile data utilizing non-linear least square fitting by internal reference spectra.

For the Auger measurements [12] a Physical Electronics PHI 660 Auger microprobe was used. The PHI 660 microprobe, an instrument with a $LaB_6$ electron emitter, has a lateral resolution of ~100 ... 200 nm under analytical working conditions using a reasonable high primary beam current. The PHI 660 microprobe is equipped with a differentially pumped $Ar^+$ ion sputter gun. In the measurement position the sample is inclined by 30° with respect to the axis of the cylindrical mirror energy analyser. Under this working condition the $Ar^+$ ions impinge on the surface at an angle of 55° with respect to the surface normal. The Unix software PHI-Access was utilized for data evaluation [13].

The sputter rates of both instruments were calibrated using a 104.6 nm $SiO_2$ layer on a Si substrate as reference material [14]. The $SiO_2$ was thermally grown on a Si wafer. The layer thickness was evaluated by ellipsometry.

## 3. Experimental Results

The benefit of wide energy range spectrum acquisition during sputter depth profile measurement using AES and XPS is demonstrated by two examples. The first depth profile measurement analyses a contamination layer containing a lot of unknown elements, which are not predictable before the analysis. The second one is from a sample, which shows large shifts of the spectra because some layers are very insulating. One constraint recording a wide energy range spectrum has to be mentioned: Higher adsorption on the freshly sputtered surface from the residual vacuum is expected since it takes a long time to measure a wide energy range spectrum [15]. Mainly the O signal, which might be due to residual water, and the C signal, which might originate from organic compounds, should be interpreted cautiously.

### 3.1 Contamination Layer on an Al Foil

In an XPS instrument a thin Al foil separates the X-ray source from the sample and the other components. This foil hinders stray electrons to escape from the X-ray source. This way electron induced sample modifications and an influence to the electron spectrometer are avoided. During sputter depth profiling the sputtered material is emitted into the half-sphere above the sample and thus it is deposited everywhere on the inner surfaces of the instrument. This way the Al foil becomes contaminated [3]. It has to be replaced from time to time, because the contamination layer reduces the X-ray flux. The drawing in the upper part of fig. 1 shows the geometrical situation in an XPS microprobe Quantum 2000. Just above the sample the Al foil window of the X-ray source and its mechanical mounting, parts of the $Ar^+$ ion gun and the electron energy analyser entrance are mounted. Together with the electron neutraliser these devices are fixed to a solid metal block. The lower part of fig. 1 shows an Al foil, which has been in use for some years. Within the exposed area the colour has changed. This contamination layer was analysed by an XPS depth profile measurement. $Ar^+$ ions of 2 keV were used for sputter erosion. The results of this depth profile measurement are depicted in fig. 2. A plot of all wide energy range XPS spectra is shown at the bottom. The XPS signal intensity is plotted against the binding energy and the sputter depth. The XPS intensity is coded by colours on a linear scale. The upper part of the fig. 2 shows a sum of all measured XPS spectra. The XPS signal is plotted against the binding energy. By means of this spectrum the elements are identified, which are present in the contamination layer. With this knowledge the measured data are re-evaluated.

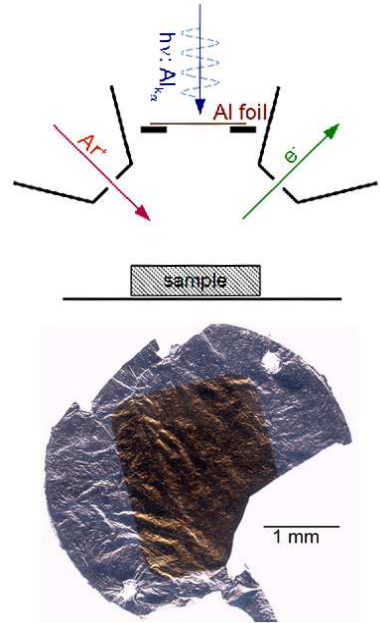

Fig. 1: top: principle drawing of the Quantum 2000 sample region
bottom: used Al foil applied as separation between X-ray source and sample and whole rest of the XPS instrument, respectively

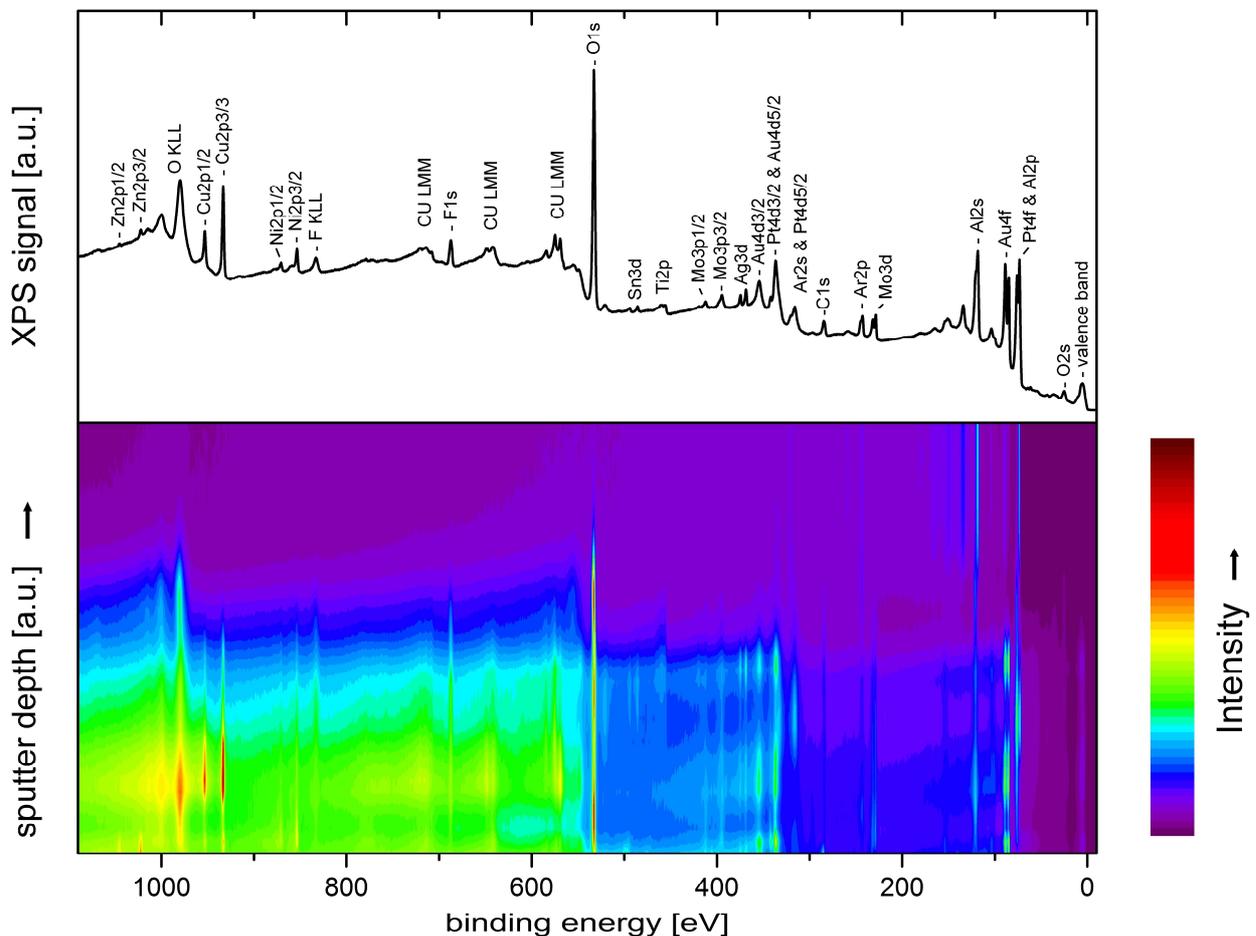

Fig. 2: top: sum of all measured XPS spectra of all sputter depths; identified peaks are labelled
bottom: XPS signal intensity as function of binding energy and sputter depth
The XPS intensity is coded by colours on a linear scale.

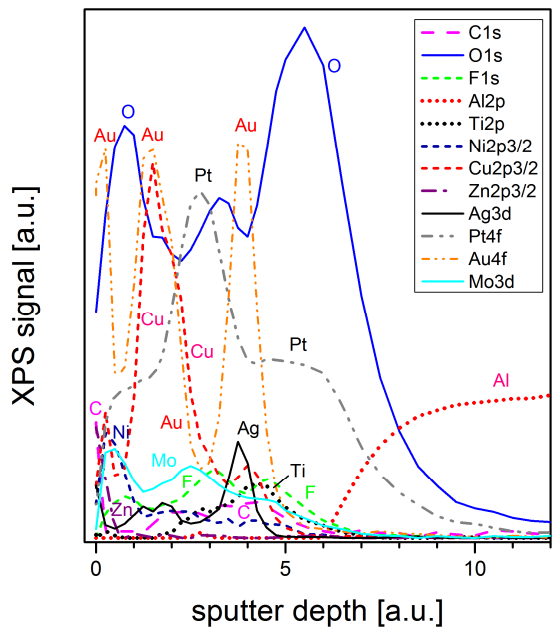

Fig. 3: sputter depth profile of the contamination layer on the Al foil

Fig. 3 shows the depth distribution of the detected elements. These elements and their depth distribution record the history of past sputter depth profile measurements. The contamination layer is estimated to be ~35 nm $SiO_2$ equivalents thick. Most likely the geometrical thickness of the layer is larger since the sputter rates of metals are higher than the $SiO_2$ sputter rate [16, 17].

## 3.2 Auger Analysis of an Insulating Thin Film System

Anticipating the results of the depth profile measurement (fig. 6) an Au / Ta / Si-C / Si-N / Al thin film system has to be analysed utilizing an Auger microprobe. The measurements were done with a 10 keV primary electron beam. For sputtering 3 keV $Ar^+$ ions were applied. Fig. 4 shows a wide energy range spectrum recorded in a sputter depth of ~3.2 µm $SiO_2$ equivalents. For lower kinetic energies up to ~230 eV the spectrum shows behaviour, which is typical for sample charging. In the direct spectrum the electron detector is overloaded. Therefore the differentiated spectrum is irregular shaped in this energy region. Auger signals are detected at ~505 eV, ~1526 eV and ~1742 eV. The peaks are not identified by their energy position but utilizing the energy difference between them. The peaks are identified this way and by the peak shapes as N, Al and Si. The whole spectrum is shifted to higher kinetic energy by ΔE ~130 eV. The whole depth profile was measured using wide energy range spectra because of this unpredictable and relatively large energy shifting. Fig. 5 shows some of these spectra exemplarily. The differentiated Auger signals are plotted against the kinetic energy and the sputter depth. The detected elements were identified using peak energy differences and peak shapes. Some of the spectra were energy shifted to lower energies prior plotting to compensate the sample charging. This energy shift is given in the drawing. Peak shifts to higher energies up to ΔE ~140 eV are observed. The disturbed low energy regions of the shifted spectra are not plotted. Since now the correct peak identification becomes possible, the depth profile can be re-evaluated. Fig. 6 shows the re-evaluated depth profile. The peak intensities were quantified using the tabulated sensitivity factors of the PHI-Access software [13].

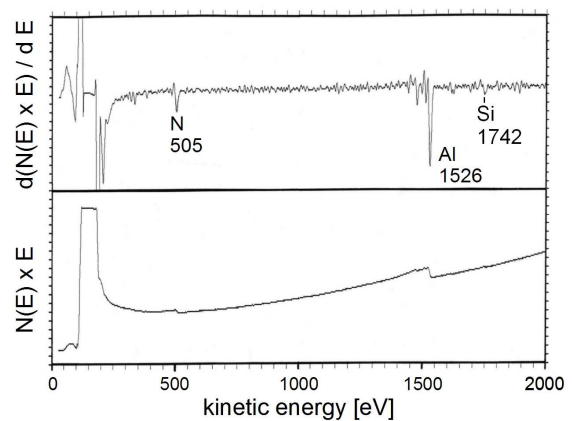

Fig. 4: shifted AES spectrum, sputter depth: 3.2 µm SiO2 equivalents

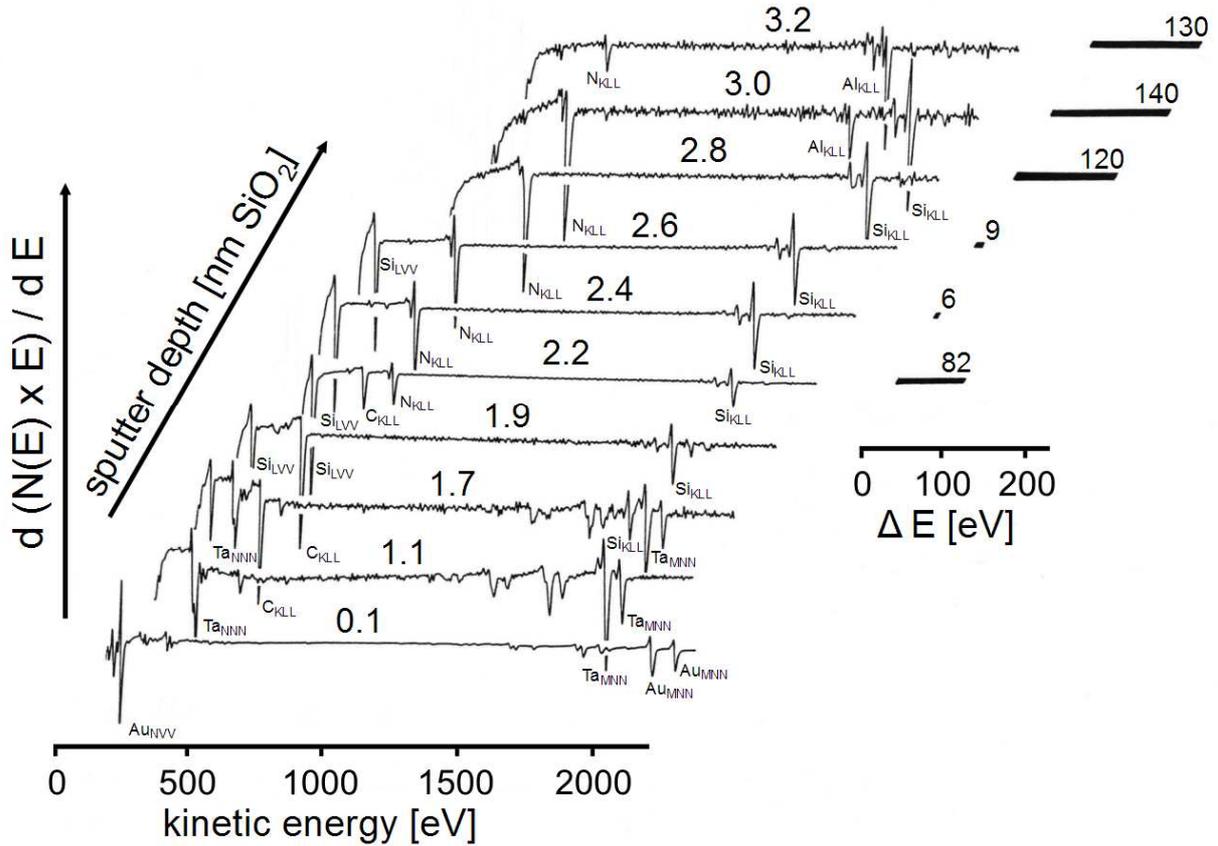

Fig. 5: 3d view of selected wide energy range spectra measured in different sputter depths
The spectra were shifted after the measurement. The energy shift ΔE of the spectra is given.

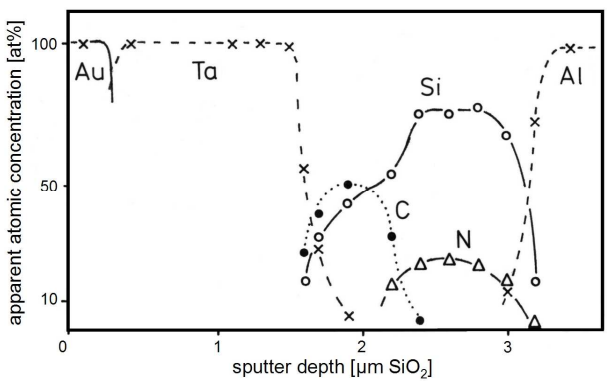

Fig. 6: re-evaluated sputter depth profile, detected layers: Au / Ta / Si-C / Si-N / Al

## 4. Conclusions

The benefit of wide energy range spectra acquisition during a sputter depth profile measurement was demonstrated exemplarily by the analyses of a contamination layer, which has a completely unpredictable composition, and for a thin film system sample, which shows large and unpredictable energy shifts of the spectra due to very insulating layers. The recording of wide energy range spectra measures the peaks of all elements, which are detectable by the used analytical method. The whole data set can be re-evaluated after the present elements are identified. In case of large energy scale shifts due to sample charging these shifts can be corrected before data re-evaluation.

The measurement of wide energy range spectra takes a long time in each sputter depth. Recontamination of the sample surface from the vacuum is expected. Therefore O and C signals should be interpreted with some precaution. Alternatively the O and C signal could be measured using small energy windows before the wide energy range spectrum is recorded in a certain sputter depth.

## Acknowledgement

The measurements were done utilizing an XPS microprobe Quantum 2000 and an Auger microprobe PHI 660, respectively, installed at Siemens AG, Munich, Germany. I acknowledge the permission of the Siemens AG to use the measurement results here. For fruitful discussions and suggestions I would like to express my thanks to my colleagues. Special thanks also to Gabi for text editing.

___